\documentclass[preprint,12pt]{elsarticle}
\usepackage{graphicx}
\usepackage{subfigure}
\usepackage{amssymb}
\usepackage{amsmath}
\usepackage{amsthm}
\usepackage{color}
\usepackage{amsfonts}
\usepackage{eepic}
\usepackage{amscd}
\usepackage{pifont,natbib,geometry,fleqn,txfonts}

\newtheorem{theorem}{Theorem}[section]
\newtheorem{proposition}{Proposition}[section]
\newtheorem{lemma}{Lemma}[section]

\begin{document}

\title{Asymptotic behaviour of a conservative reaction-diffusion system associated with a Markov process algebra model}

\author[jd1]{Jie~Ding}
\ead{jieding78@hotmail.com}

\author[ma]{Runmin Ma}
\ead{yzmrm0107@gmail.com }

\author[yz]{Zhigui~Lin}
\ead{zglin68@hotmail.com}

\author[yz]{Zhi~Ling}
\ead{zhling@yzu.edu.cn 
}

\address[jd1]{School of Computer, Jiangsu University of Science and Technology, Zhenjiang 212100, China}

\address[ma]{School of Computer Science and Engineering, Northeastern University, Shenyang 110000, China}

\address[yz]{School of Mathematics, Yangzhou University,
Yangzhou, 225002, China}

\begin{abstract}
This paper  demonstrates a lower and upper solution method to
investigate the asymptotic behaviour of the conservative
reaction-diffusion systems   associated with  Markovian process algebra models.
In particular, we have proved the uniform convergence of the solution
to its constant equilibrium for a case study as time tends to infinity,
together with experimental results illustrations.

\end{abstract}


\maketitle

\section{Introduction}

Markovian process algebras like PEPA~\cite{Jane1}
are powerful formal tools for performance modelling of concurrent computer and communication  systems~\cite{HaoWang-IEEE-Mobile-Computing2013},
supply chains~\cite{Jie-CAIE2022} and block-chains~\cite{Jie-IEEE-ITS2022},
as well as biochemical networks~\cite{Bio-PEPA} and
epidemiological systems~\cite{SB-Improved-ContinuousApproximation-Epidemiological, Measles-Epidemics-PEPA2012}.
However, such discrete state-based modelling formalisms
are challenged by the size and complexity of large scale systems, i.e., there is a
state-space explosion problem encountered
in calculating the steady-state probability distributions of the underling Markov chains.
Fluid approximation approaches have been proposed to
to deal with this problem, which utilises a set of ordinary differential
equations (ODEs) to approximate the underling continuous-time
Markov chains (CTMCs)~\cite{Jane2, Mirco-TSE2012, Hayden-TSE2013}.
Nevertheless, geographical information sometimes can not be neglected in
modelling a mobile current system such like  collective robots systems, self-driving vehicle networks,  etc.
Therefore, the ODEs are  extended to partial differential equations (PDEs) to incorporate
spatial content. In the PDEs context, it is   the
evolution of the densities rather than the populations of the entities
to be considered, with an emphasis on the effect of dispersion in a bounded
region, and in this situation the governing equations for the
population densities are described by a system of
reaction-diffusion equations~\cite{Jie-AML2011, Jie-AMM2013}.
In fact, the links between reaction-diffusions and the CTMCs
underlying PEPA models have been  revealed~\cite{PEPA:PDE2CTMC} for a biochemical system in pioneering.

\par The asymptotic behaviour of the ODEs derived through fluid approximation, and the relationship with
the underlying CTMCs of Markovian process algebra models, has been
intensively investigated theoretically or experimentally~\cite{JieThesis, Mirco-TSE2012, Hayden-TSE2013}.
However, not much work relates to the long-time behaviour of the associated reaction-diffusion systems,
except for~\cite{Jie-AML2011, Jie-AMM2013}.  One reason is that the PDEs  cannot be reduced to ordinary differential equations  to treat as a usual practice dealing with  null Nuemann boundary conditions, because the involved ``minimum'' functions which
are determined by the operational semantics of Markovian process algebras  are not differentiable.

In this paper, we will provide
an upper and lower method~\cite{Smoller-Shock-Wave-RD, CVPao-book-PDE} to approximate the conservative reaction-diffusion system.
Usually, it is difficult to find appropriate upper and lower solutions to iteratively  approximate a  conservative system.
Our trick   is to utilise its equilibrium to construct upper and and lower solutions, because an equilibrium solution is naturally  an upper solution as well as a lower solution.
Here is the outline of proof. First, we will determine the system has
a unique constant equilibrium. Then, by scaling the equilibrium, a pair of upper and lower  solutions are obtained for
initial iteration. Subsequently, at each step of iterations, the derived upper and lower solutions  will be shown uniformly convergent with time to constants, and these constants are getting closer  as the number of iterations increases, until they finally meet together in the sense of limit.
Therefore, the original solution of the system, which is sandwiched between the sequences of upper and lower solutions, is forced to converge to a limit.

\par In the following, we will demonstrate this lower and upper solution method to investigate the long-time behaviour in a case study. For convenience of comparison,   a set of conservative reaction-diffusion system   associated
with a  PEPA model is borrowed here, which is presented and investigated in paper~\cite{Jie-AMM2013}. That is,

\begin{equation}\label{eq:ChFA-Model-XY-PDEs}
\left\{
\begin{aligned}
\frac{\partial u_1}{\partial t}-\Delta u_1 &=-a_1\min\{u_1,v_1\}+a_2\min\{u_2,v_4\},\\
\frac{\partial u_2}{\partial t}-\Delta u_2 &=a_1\min\{u_1,v_1\}-a_2\min\{u_2,v_4\},\\
\frac{\partial v_1}{\partial t}-\Delta v_1 &=-a_1\min\{u_1,v_1\}-c_1v_1+c_2v_2,\\
\frac{\partial v_2}{\partial t}-\Delta v_2 &=-c_2v_2+c_1v_1+a_2\min\{u_2,v_4\},\\
\frac{\partial v_3}{\partial t}-\Delta v_3 &=-c_3v_3+c_4v_4+a_1\min\{u_1,v_1\},\\
\frac{\partial v_4}{\partial t}-\Delta v_4
&=-a_2\min\{u_2,v_4\}-c_4v_4+c_3v_3,
\end{aligned}
 \right.
\end{equation}
in $\Omega\times [0,\infty)$, where $\Omega\subset
\mathbb{R}^d$ and $d$ could be one, two or other positive
integers. Here $u_i(x,t), v_i(x,t)$ in
(\ref{eq:ChFA-Model-XY-PDEs}) are  the population densities of some entities
  distributed on a region $\Omega$
at time $t$. In addition,  for convenience, the diffusion
constants are set to be one in these PDEs.
In this paper we are concerned with the following boundary and
initial conditions:
\begin{equation}\label{eq:PDEs-Boundary}
       \frac{\partial u_1}{\partial\eta}=\frac{\partial u_2}{\partial\eta}=\frac{\partial
       v_1}{\partial\eta}=\frac{\partial v_2}{\partial\eta
       }=\frac{\partial v_3}{\partial\eta}=\frac{\partial v_4}{\partial\eta
       }=0,
\end{equation}
\begin{equation}\label{eq:PDEs-Initial}
u_i(x,0)=\phi_i(x),\ v_j(x,0)=\psi_j(x),\  i=1,2,\ j=1,2,3,4.
\end{equation}
Here $\eta$ is an outer normal vector along the
boundary of $\Omega$. The null Neumann boundary condition
(\ref{eq:PDEs-Boundary}) means that there is no immigration across
the boundary. The initial distributions of $u_i$ and $v_i$ on
$\Omega$ are given in the initial condition
(\ref{eq:PDEs-Initial}).

\par For detailed introduction to the PDEs and their background, please see  paper~\cite{Jie-AMM2013}. The following results regarding the existence,  boundedness and positivity of solution,  have been established~\cite{Jie-AMM2013}.
\begin{theorem}(\cite{Jie-AMM2013})
The solution of  system (\ref{eq:ChFA-Model-XY-PDEs}) with
boundary condition (\ref{eq:PDEs-Boundary}) and initial condition
(\ref{eq:PDEs-Initial}) globally exists in $[0, \infty)$.
\end{theorem}

\begin{theorem}\label{thm:bounded-solution}(\cite{Jie-AMM2013})
 Let $(u_1, u_2, v_1,v_2,v_3,v_4)$ be the solution of system
(\ref{eq:ChFA-Model-XY-PDEs}) with the boundary condition
(\ref{eq:PDEs-Boundary}) and initial condition
(\ref{eq:PDEs-Initial}). Suppose that $\phi_i$ and $\psi_j$ are
positive,  $i=1,2,\ j=1,2,3,4$. Then the solution is uniformly bounded in
  ${\Omega}\times(0,+\infty)$, and the solution is positive, i.e.\ $u_i>0$ and
$v_j>0$ in $\Omega\times(0,+\infty)$ for $i=1,2,\ j=1,2,3,4$.
\end{theorem}

In addition,  the solution of system
(\ref{eq:ChFA-Model-XY-PDEs})-(\ref{eq:PDEs-Initial}), has been shown  in~\cite{Jie-AMM2013}
to  converge to some constants as time goes to infinity,  under some  conditions on the parameters and entity populations. The methods of proof presented in~\cite{Jie-AMM2013} are essentially relied on the conditions and hence they are unremovable.
Further, even in some particular situations, such similar conditions are not been found so that only numerical experiments without theoretical results are utilised to demonstrate the convergence in~\cite{Jie-AMM2013}.

In contrast, as a main contribution of this paper, we will demonstrate the following convergence result  without any condition.
\begin{theorem}\label{thm:main-convergence}
The solution of the system (\ref{eq:ChFA-Model-XY-PDEs}) with the
boundary condition (\ref{eq:PDEs-Boundary}) and initial condition
(\ref{eq:PDEs-Initial}) uniformly converges to its unique constant equilibrium as time tends to infinity.
\end{theorem}

\par  Before we give a complete proof to this theorem, we should point out that
if   Dirichlet boundary conditions are considered instead, then we also have a similar asymptotic
conclusion:
 \begin{theorem}\label{thm:Dirichelet-main-convergence}
  System (\ref{eq:ChFA-Model-XY-PDEs}) with the
initial condition (\ref{eq:PDEs-Initial}) and the following homogenous Dirichlet boundary condition
 $$
     u_1(x,t)=u_2(x,t)=v_1(x,t)=v_2(x,t)=v_3(x,t)=v_4(x,t)=0, \quad x\in\partial\Omega,
 $$
has a unique solution which converges to zeros uniformly as time tends to infinity.
\end{theorem}
The proof is simple and we only  sketch it here.  The first step is to
determine that the solution is nonnegative, which can be proved similarly to the case
of Nuemann boundary condition~\cite{Jie-AML2011,Jie-AMM2013}. Subsequently,  consider $h(x,t)=\sum_{i=1}^2u_i(x,t)+\sum_{j=1}^4v_j(x,t)$, which satisfies that
\begin{equation*}
\left\{
\begin{array}{ll}
  \frac{\partial h}{\partial t}-\Delta h=0, & (x,t)\in\Omega\times(0,+\infty), \\
   h(x,t)=0,& (x,t)\in\partial\Omega\times(0,+\infty),\\
  h(x,0)=\sum_{i=1}^2\phi_i(x)+\sum_{j=1}^4\psi_j(x), &x\in\Omega.
\end{array}
\right.
\end{equation*}
It is well known that $h(x,t)$ uniformly tends to zero as time goes to infinity. Because
all $u_i(x,t) (i=1,2)$ and $v_j(x,t)(j=1,2,3,4)$ are nonnegative, they consequently converges to zeros uniformly.
All rest work in the paper is to turn to prove Theorem~\ref{thm:main-convergence}.

\section{Constant equilibrium exists and is unique}

 We   rewrite system  (\ref{eq:ChFA-Model-XY-PDEs}) as follows:
\begin{equation}\label{eq:ChFA-Model-XY-PDEQ1Q2Q3Q4}
\begin{aligned}
&(\frac{\mathrm{\partial}u_1}{\mathrm{\partial}t},\frac{\mathrm{\partial}u_2}{\mathrm{\partial}t},\frac{\mathrm{\partial}v_1}{\mathrm{\partial}t},
\frac{\mathrm{\partial}v_2}{\mathrm{\partial}t},\frac{\mathrm{\partial}v_3}{\mathrm{\partial}t},\frac{\mathrm{\partial}v_4}{\mathrm{\partial}t})^{T}
-(\Delta u_1,\Delta u_2,\Delta v_1,\Delta v_2,\Delta v_3,\Delta
v_4)^{T}
\\=&I_{\{ u_1<v_1,
                u_2< v_4
           \}}Q_1(u_1,u_2,v_1,v_2,v_3,v_4)^{T}
 +I_{\{u_1<v_1, u_2\geq v_4\}}Q_2(u_1,u_2,v_1,v_2,v_3,v_4)^{T}
\\
&+I_{\{ u_1\geq v_1,
                u_2<v_4
           \}}Q_3(u_1,u_2,v_1,v_2,v_3,v_4)^{T}
 +I_{\{u_1\geq v_1, u_2\geq v_4\}}Q_4(u_1,u_2,v_1,v_2,v_3,v_4)^{T},
\end{aligned}
\end{equation}
where $I$ is an indicator function, and the matrices
$Q_i\;(i=1,2,3,4)$ are given as below:
\begin{eqnarray*}
Q_1=\left(
  \begin{array}{cc|cccc}
   -a_1 & a_2 &  &  &  &  \\
    a_1 &  -a_2  &  &  &  \\\hline
    -a_1 &  & -c_1 & c_2 &  &  \\
     & a_2 &  c_1 & -c_2 &  &  \\
    a_1 &  &  &  & -c_3 & c_4 \\
     & -a_2 &  &  & c_3 & -c_4
  \end{array}
\right),
\end{eqnarray*}
\begin{eqnarray*}
Q_2=\left(
  \begin{array}{cc|cccc}
   -a_1 & 0 &  &  &  & a_2 \\
    a_1 & 0 &  &  &  & -a_2 \\\hline
    -a_1 & 0 & -c_1 & c_2 &  &  \\
     & 0 &  c_1 & -c_2 &  & a_2 \\
    a_1 & 0 &  &  & -c_3 & c_4 \\
     & 0 &  &  & c_3 & -(c_4+a_2)
  \end{array}
\right),
\end{eqnarray*}
\begin{eqnarray*}
Q_3=\left(
  \begin{array}{cc|cccc}
   0& a_2 &  -a_1 &  &  &  \\
   0 &  -a_2 & a_1  &  &  &  \\\hline
   0  &  & -a_1-c_1 & c_2 &  &  \\
   0  & a_2 &  c_1 & -c_2 &  &  \\
   0  &  & a_1 &  & -c_3 & c_4 \\
   0  & -a_2 &  &  & c_3 & -c_4
  \end{array}
\right),
\end{eqnarray*}
\begin{eqnarray*}
Q_4=\left(
  \begin{array}{cc|cccc}
   0 & 0 & -a_1 &  &  & a_2 \\
   0 &  0 & a_1 &  &  &-a_2  \\\hline
    0 & 0 & -a_1-c_1 & c_2 &  &  \\
    0 & 0 &  c_1 & -c_2 &  & a_2 \\
    0 & 0 &  a_1  && -c_3 & c_4 \\
    0 &0 &  &  & c_3 & -a_2-c_4
  \end{array}
\right).
\end{eqnarray*}

The solution of system~(\ref{eq:ChFA-Model-XY-PDEs})-(\ref{eq:PDEs-Initial})  satisfies a conservation law, which
is specified in the following lemma.
\begin{lemma}\label{lemma:lemma-conservation2}(\cite{Jie-AMM2013})
As time goes to infinity, the solution of system~(\ref{eq:ChFA-Model-XY-PDEs})-(\ref{eq:PDEs-Initial}) satisfies that
as time $t$ tends to infinity,
    \begin{equation}
    \sum^{2}_{i=1}{u_i(x,t)}\rightrightarrows M_0=\frac{1}{|\Omega|}\int_{\Omega}(\phi_1+\phi_2)dx,
    \end{equation}
\begin{equation}
 \sum^{4}_{j=1}{v_j(x,t)}\rightrightarrows N_0=\frac{1}{|\Omega|}\int_{\Omega}(\psi_1+\psi_2+\psi_3+\psi_4)dx,
\end{equation}
\begin{equation}
  (v_1+v_2-u_1)\rightrightarrows W_{1}=\frac{1}{|\Omega|}\int_{\Omega}(\psi_1+\psi_2-\phi_1)dx,
  \end{equation}
\begin{equation}
 (v_3+v_4-u_2)\rightrightarrows W_{2}=\frac{1}{|\Omega|}\int_{\Omega}(\psi_3+\psi_4-\phi_2)dx.
\end{equation}
Hereafter ``$\rightrightarrows$'' means uniform convergence in $L^2(\Omega)$, and $M_0,N_0, W_1, W_2$ are defined as the above ones.
\end{lemma}

Consider the equilibrium equation
\begin{equation}\label{eq:equilibrium-equation}
\begin{aligned}
 &I_{\{ u_1^*<v_1^*,
                u_2^*< v_4^*
           \}}Q_1(u_1^*,u_2^*,v_1^*,v_2^*,v_3^*,v_4^*)^{T}
 +I_{\{u_1^*<v_1^*, u_2^*\geq v_4^*\}}Q_2(u_1^*,u_2^*,v_1^*,v_2^*,v_3^*,v_4^*)^{T}
\\
&+I_{\{ u_1^*\geq v_1^*,
                u_2^*<v_4^*
           \}}Q_3(u_1^*,u_2^*,v_1^*,v_2^*,v_3^*,v_4^*)^{T}
 +I_{\{u_1^*\geq v_1^*, u_2^*\geq v_4^*\}}Q_4(u_1^*,u_2^*,v_1^*,v_2^*,v_3^*,v_4^*)^{T}=0,
\end{aligned}
\end{equation}
with
\begin{equation}\label{eq:equilibrium-equation-2}
\left\{
\begin{aligned}
&u_1^*+u_2^*=M_0,\\
&v_1^*+v_2^*+v_3^*+v_4^*=N_0,\\
&v_1^*+v_2^*-u_1^*=W_1, \ v_3^*+v_4^*-u_2^*=W_2.
\end{aligned}\right.
\end{equation}
We will show that equilibrium system (\ref{eq:equilibrium-equation})-(\ref{eq:equilibrium-equation-2}) and therefore system (\ref{eq:ChFA-Model-XY-PDEs})-(\ref{eq:PDEs-Initial}),
 admit a  unique constant equilibrium.
For convience, we define constants:
$$D_1=a_1c_2(a_2+c_4)+a_2c_3(a_1+c_1)+c_2c_3(a_1+a_2),$$
$$D_2=a_1(a_2+c_3+c_4)+a_2c_3,\ D_3=a_2(a_1+c_1+c_2)+a_1c_2.$$
In addition, we define conditions $(I_i), (I_i^c), i=1,2,3,4$, as follows:
\begin{equation*}\label{eq:I1}
(I_1): a_2(a_1 + c_1)M_0 < c_2(a_1 + a_2)W_1;\quad  (I_1^c): a_2(a_1 + c_1)M_0 \geq c_2(a_1 + a_2)W_1;
\end{equation*}
\begin{equation*}\label{eq:I2}
(I_2): a_1(a_2 + c_4)M_0 < c_3(a_1 + a_2)W_2; \quad (I_2^c): a_1(a_2 + c_4)M_0 \geq  c_3(a_1 + a_2)W_2;
\end{equation*}
\begin{equation*}\label{eq:I3}
(I_3): a_2c_3(a_1 + c_1)N_0 < D_1W_1; \quad (I_3^c): a_2c_3(a_1 + c_1)N_0 \geq  D_1W_1;
\end{equation*}
\begin{equation*}\label{eq:I4}
(I_4): a_1c_2(a_2 + c_4)N_0 < D_1W_2; \quad  (I_4^c): a_1c_2(a_2 + c_4)N_0 \geq  D_1W_2.
\end{equation*}

\begin{lemma}\label{lemma:equibrium}
Suppose the initial condition (\ref{eq:PDEs-Initial}) is positive.
\begin{enumerate}

  \item ($I_1\wedge I_2$: $Q_1$ dominates) If conditions $(I_1)$ and $(I_2)$ are satisfied, then the unique equilibrium is:\\
   $ (\frac{M_0a_2}{a_1+a_2},
   \frac{M_0a_1}{a_1+a_2},
   \frac{a_2(c_2-a_1)M_0 + c_2(a_1+a_2)W_1}{(a_1 + a_2)(c_1 + c_2)},
   \frac{a_2(a_1+ c_1)M_0 + c_1(a_1+a_2)W_1}{(a_1 + a_2)(c_1 + c_2)},
    \frac{a_2(a_1-c_4)M_0 +c_4(a_1+a_2)(N_0-W_1) }{(a_1+a_2)(c_3+c_4)},\\
   \frac{-a_2(a_1+c_3)M_0+c_3(a_1+a_2)(N_0-W_1)}{(a_1+a_2)(c_3+c_4)} ).$

\item ($I_3\wedge I_2^c$: $Q_2$ dominates) If conditions $(I_3)$ and $(I_2^c)$ are satisfied,  then the unique equilibrium is:\\
$ (\frac{a_2c_3}{D_2}(M_0+W_2),(1-\frac{a_2c_3}{D_2})M_0-\frac{a_2c_3}{D_2}W_2,
\frac{c_2}{c_1+c_2}N_0-\frac{a_1c_2(a_2+c_3+c_4)+a_1a_2c_3}{D_2(c_1+c_2)}(M_0+W_2),
\frac{c_1}{c_1+c_2}N_0-\frac{a_1c_1(a_2+c_3+c_4)-a_1a_2c_3}{D_2(c_1+c_2)}(M_0+W_2),
\frac{a_1(a_2+c_4)}{D_2}(M_0+W_2),\frac{a_1c_3}{D_2}(M_0+W_2) )^{T}$.

\item ($I_4\wedge I_1^c$: $Q_3$ dominates) If conditions $(I_4)$ and $(I_1^c)$ are satisfied, then the unique equilibrium is:\\
$((1-\frac{a_1c_2}{D_3})M_0-\frac{a_1c_2}{D_3}W_1,
\frac{a_1c_2}{D_3}(M_0+W_1),\frac{a_2c_2}{D_3}(M_0+W_1),\frac{a_2(a_1+c_1)}{D_3}(M_0+W_1),
\frac{c_4}{c_3+c_4}N_0-\frac{a_2c_4(a_1+c_1+c_2)-a_1a_2c_2}{D_3(c_3+c_4)}(M_0+W_1),
  \frac{c_3}{c_3+c_4}N_0-\frac{a_2c_3(a_1+c_1+c_2)+a_1a_2c_2}{D_3(c_3+c_4)}(M_0+W_1))^{T}$.

  \item ($I_3^c\wedge I_4^c$: $Q_4$ dominates) If conditions $(I_3^c)$ and $(I_4^c)$ are satisfied, then the unique equilibrium is:\\
  $ (\frac{a_2c_3(a_1+c_1+c_2)}{D_1}N_0-W_1,
M_0+W_1-\frac{a_2c_3(a_1+c_1+c_2)}{D_1}N_0,  \frac{a_2c_2c_3}{D_1}N_0,
  \frac{a_2c_3(a_1+c_1)}{D_1}N_0,
\frac{a_1c_2(a_2+c_4)}{D_1}N_0,\frac{a_1c_2c_3}{D_1}N_0 )^{T}$.

\end{enumerate}
Hereafter, ``$Q_i$ dominates'' is refered to as $Q_i(u_1^*,u_2^*,v_1^*,v_2^*,v_3^*,v_4^*)^T=0$.
\end{lemma}

\begin{proof} We only prove the first term. Let
   $$u_1^*=\frac{M_0a_2}{a_1+a_2}, \ u_2^*=\frac{M_0a_1}{a_1+a_2}, $$
  $$ v_1^*=\frac{a_2(c_2-a_1)M_0 + c_2(a_1+a_2)W_1}{(a_1 + a_2)(c_1 + c_2)}, \
  v_2^*= \frac{a_2(a_1+ c_1)M_0 + c_1(a_1+a_2)W_1}{(a_1 + a_2)(c_1 + c_2)},$$
   $$ v_3^*=\frac{a_2(a_1-c_4)M_0 +c_4(a_1+a_2)(N_0-W_1) }{(a_1+a_2)(c_3+c_4)},  \
  v_4^*= \frac{-a_2(a_1+c_3)M_0+c_3(a_1+a_2)(N_0-W_1)}{(a_1+a_2)(c_3+c_4)}.$$
Clearly, condition $(I_1)$ is equivalent to $u_1^*<v_1^*$, and condition $(I_2)$ is equivalent to $u_2^*<v_4^*$.
Then, it is easy to verify that  $Q_1(u_1^*,u_2^*,v_1^*,v_2^*,v_3^*,v_4^*)^T=0$, i.e.,
 $(u_1^*,u_2^*,v_1^*,\cdots, v_4^*)$ satisfies (\ref{eq:equilibrium-equation}) and (\ref{eq:equilibrium-equation-2}).
By checking the rank of $Q_1$ with (\ref{eq:equilibrium-equation-2}), the uniqueness is also clear.

\end{proof}

\begin{proposition}
For all positive parameters and positive initial functions,
system (\ref{eq:ChFA-Model-XY-PDEs})-(\ref{eq:PDEs-Initial}) always exists a unique constant equilibrium.
\end{proposition}
\begin{proof}
By $\mathcal{S} $ we denote the set of all parameters   and positive initial functions, i.e.,
$$
  \mathcal{S} =\left\{ \left(a_1,a_2, c_1,\cdots, c_4;\phi_1,\phi_2, \psi_1,\cdots,\psi_4\right )\right\}.
$$
Let $\mathcal{S}_{ A }$ indicate  the subset of $\mathcal{S}$ whose elements satisfy condition $ A $, i.e.,
$$\mathcal{S}_{ A }=\{\omega\in \mathcal{S}\mid \omega \  \mathrm{ satisfies } \  A \}.$$
The proposition is equivalent to that
$$\mathcal{S} = \mathcal{S}_{ I_1\wedge I_2 }\cup \mathcal{S}_{I_3\wedge I_2^c} \cup \mathcal{S}_{I_4\wedge I_1^c} \cup \mathcal{S}_{I_3^c\wedge I_4^c}.$$
We only need to prove
$\mathcal{S} \subseteq \mathcal{S}_{ I_1\wedge I_2 }\cup \mathcal{S}_{I_3\wedge I_2^c} \cup \mathcal{S}_{I_4\wedge I_1^c} \cup \mathcal{S}_{I_3^c\wedge I_4^c}$.
Suppose $\omega\in\mathcal{S}$
but  $\omega\notin \mathcal{S}_{I_1\wedge I_2}\cup \mathcal{S}_{I_3\wedge I_2^c} \cup \mathcal{S}_{I_4\wedge I_1^c} \cup \mathcal{S}_{I_3^c\wedge I_4^c}$. That is to say, $\omega\in \left(\mathcal{S}_{I_1\wedge I_2}\cup \mathcal{S}_{I_3\wedge I_2^c} \cup \mathcal{S}_{I_4\wedge I_1^c} \cup \mathcal{S}_{I_3^c\wedge I_4^c}\right)^c$.
This implies that
$\omega\in\mathcal{S}_{I_1\wedge I_2^c \wedge I_3^c \wedge I_4}$ or
$\omega\in\mathcal{S}_{I_1^c\wedge I_2 \wedge I_3 \wedge I_4^c}$. We will show that both
$\mathcal{S}_{I_1\wedge I_2^c \wedge I_3^c \wedge I_4}$ and $\mathcal{S}_{I_1^c\wedge I_2 \wedge I_3 \wedge I_4^c}$ are empty sets.

\par In fact, according to condition  $I_1\wedge I_2^c \wedge I_3^c \wedge I_4$, we can deduce that
 \begin{equation}\label{N0}
		 \frac{D_1 W_1}{a_1 a_2 c_3+a_2 c_1 c_3}\leqslant N_0<\frac{D_1 W_2}{a_1 a_2 c_2+a_1 c_2 c_4},
 \end{equation}
 \begin{equation}\label{M_0_0}
  \frac{\left(a_1 c_3+a_2 c_3\right) W_2}{a_1 a_2+a_1 c_4}\leqslant M_0<\frac{\left(a_1 c_2+a_2 c_2\right) W_1}{a_1 a_2+a_2 c_1}.
	\end{equation}
Notice that  $N_0 = M_0 + W_1 + W_2$, then (\ref{N0}) implies that
\begin{equation}\label{M_0_1}
	\frac{D_1 - a_1 a_2 c_3-a_2 c_1 c_3}{a_1 a_2 c_3+a_2 c_1 c_3}W_1 - W_2\leqslant M_0< \frac{ D_1 - a_1 a_2 c_2-a_1 c_2 c_4 }{a_1 a_2 c_2+a_1 c_2 c_4}W_2 - W_1,
\end{equation}
which leads to
\begin{equation}\label{contradiction_2}
	W_2 >  \frac{a_1c_2(a_2 + c_4)}{a_2c_3(a_1 + c_1)}W_1.
\end{equation}
However, (\ref{M_0_0}) implies that
\begin{equation}\label{contradiction_1}
	W_2 < \frac{\left(a_1 c_2+a_2 c_2\right)(a_1 a_2 + a_1 c_4) }{(a_1 a_2+a_2 c_1)(a_1 c_3 + a_2 c_3)}W_1 = \frac{a_1c_2(a_2 + c_4)}{a_2c_3(a_1 + c_1)}W_1.
\end{equation}
This is a contradiction. Therefore, $\mathcal{S}_{I_1\wedge I_2^c \wedge I_3^c \wedge I_4}=\emptyset$. Similarly, we can   prove
$\mathcal{S}_{I_1^c\wedge I_2 \wedge I_3 \wedge I_4^c}$ is also empty.
\end{proof}

\section{Proof of convergence result}

\subsection{Preliminary}

Some lemmas are presented in this subsection,  which will be
utilised  to prove the  long-time behaviour of the system.

\begin{lemma}\label{lemma:convergence-1}
Let $|\Omega|$ be the
measure of   region $\Omega$.
\begin{enumerate}
\item If $z(x,t)$ satisfies
\begin{equation}\label{eq:lemma-conservation-1}
\left\{
\begin{array}{ll}
  \frac{\partial z}{\partial t}-\Delta z=\rho(x,t), & (x,t)\in\Omega\times(0,+\infty), \\
  \frac{\partial z}{\partial {\eta}}=0,& (x,t)\in\partial\Omega\times(0,+\infty),\\
  z(x,0)=\phi(x), &x\in\Omega,
\end{array}
\right. \end{equation}
 where $\rho(x,t)$ tends to zero uniformly as time goes to infinity,    then $z(x,t)$ uniformly converges to  $\frac{1}{|\Omega|}\int_{\Omega}\phi dx$ as $t\rightarrow \infty$.

\item If $z$ satisfies
\begin{equation}\label{eq:conservation-2}
\left\{
\begin{array}{ll}
  \frac{\partial z}{\partial t}-\Delta z=-\alpha z+\beta+\rho(x,t), &  (x,t)\in\Omega\times(0,+\infty),\\
   \frac{\partial z}{\partial \eta}=0,& (x,t)\in\partial\Omega\times(0,+\infty), \\
  z(x,0)=\phi(x),&x\in\Omega,
\end{array}
\right.
\end{equation}
where $\alpha>0$ and $\rho(x,t)$ tends to zero uniformly as $t\rightarrow \infty$,    then $z(x,t)$ uniformly converges to  $\frac{\beta}{\alpha}$ as time tends to infinity.
\end{enumerate}
\end{lemma}

\begin{proof} We prove the first conclusion. Let $0=\lambda_0<\lambda_1\leq \lambda_2\leq\cdots$ be the eigenvalues of operator $-\Delta$ with the homogenous Neumann boundary condition, and $\{\psi_k\}_{k=0}^{\infty}$ be the corresponding normal orthogonal eigenfunctions which is a set of basis of $L^2(\Omega)$. Denote
$$
    z(x,t)=\sum_{k=0}^{\infty}z_k(t)\psi_k(x),\quad  \rho(x,t)=\sum_{k=0}^{\infty}\rho_k(t)\psi_k(x),
$$
then $-\Delta z(x,t)=\sum_{k=0}^{\infty}\lambda_k z_k(t)\psi_k(x)$, with
$z_0=\frac{1}{|\Omega|}\int_\Omega \phi$. Notice that
$$|\rho_k(t)|=\left|\int_{\Omega}\rho(x,t)\psi_k(x)\right| \rightrightarrows 0,  t\rightarrow \infty.$$
This implies that for $k\geq 1$, the solution of
$$
   \frac{ \partial {z_k(t)}}{\partial t}-\lambda_k z_k(t)=\rho_k(t)
$$
converges to zero uniformly, i.e., $z_k(t)\rightrightarrows 0, t\rightarrow\infty$. Therefore,
$z(x,t)$ uniformly converges to $z_0=\frac{1}{|\Omega|}\int_\Omega \phi$, as time tends to infinity.
\par  Applying this result to
$e^{\alpha t}(z-\frac{\beta}{\alpha})$, we can obtain the second conclusion.

\end{proof}

\begin{lemma}\label{lemma:convergence-2}  Let $c>0$, $a, b,d\geq 0$,  and $w(x,t)$ satisfy
\begin{equation}\label{eq:lemma-convergence-2}
\left\{
\begin{array}{ll}
  \frac{\partial w}{\partial t}-\Delta w=-a\min\{w, b\}-cw+d+\rho(x,t), & (x,t)\in\Omega\times(0,+\infty), \\
  \frac{\partial w}{\partial {\eta}}=0,& (x,t)\in\partial\Omega\times(0,+\infty),\\
  w(x,0)=\phi(x)> 0, &x\in\Omega.
\end{array}
\right. \end{equation}
Here  $ \rho(x,t) $ uniformly converges to zero as time goes to infinity. Then as time goes to infinity, $w(x,t)$ uniformly converges to  $\frac{d}{a+c}$ if
$\frac{d}{a+c}\leq b$, or otherwise to $\frac{d-ab}{c}$.
\end{lemma}

\begin{proof}
Noticing
$$
    \frac{\partial w}{\partial t}-\Delta w\geq-aw-cw+d+\rho(x,t),
$$
by the comparison principle and Lemma~\ref{lemma:convergence-1}, we know that for any $\epsilon>0$ there exists $T>0$ such that
$\forall (x,t)\in \Omega\times (T,+\infty), w(x,t)\geq \frac{d}{a+c}-\epsilon.$

If $ \frac{d}{a+c}>b $, choose $\epsilon<|\frac{d}{a+c}-b|$, then after time $T$,
$$
 \frac{\partial w}{\partial t}-\Delta w=-a\min\{w, b\}-cw+d+\rho(x,t) =-ab-cw+d+\rho(x,t),
$$
which results in that  $w(x,t)\rightrightarrows \frac{d-ab}{c}, t\rightarrow \infty$, according to Lemma~\ref{lemma:convergence-1}.

\par
Otherwise, if $\frac{d}{a+c}\leq b $, by noticing that after time $T$,   $w(x,t)>\frac{d}{a+c}-\epsilon$, so
\begin{equation*}
\begin{aligned}
 \frac{\partial w}{\partial t}-\Delta w &=-a\min\{w, b\}-cw+d+\rho(x,t) \\
 &\leq-a\min\left\{\frac{d}{a+c}-\epsilon, b\right\}-cw+d+\rho(x,t)\\
 &=-a\frac{d}{a+c}+a\epsilon-cw+d+\rho(x,t).
\end{aligned}
\end{equation*}
Consequently, according to the comparison principle and Lemma~\ref{lemma:convergence-1}, we   know that there exists $T'>T>0$ such that $w(x,t)\leq \frac{d}{a+c}+\epsilon$, and therefore, after time $T'$,
$$
  \frac{d}{a+c}-\epsilon \leq  w(x,t)\leq \frac{d}{a+c}+\epsilon.
$$
That is, $w(x,t)\rightrightarrows \frac{d}{a+c}, t\rightarrow \infty$.
\end{proof}

\begin{lemma}\label{lemma:lemma-conservation3} (~\cite{Jie-AMM2013})
Let $A$ be a real matrix with all eigenvalues being either zeros
or having negative real parts. If $\mathbf{z}=(z_1,z_2,\cdots,z_n)^{T}$ is
uniformly bounded  in $\Omega\times(0,+\infty)$,  and satisfies
\begin{equation}\label{eq:lemma-conservation-3}
\left\{
\begin{array}{ll}
  \frac{\partial \mathbf{z}}{\partial t}-\Delta U=A\mathbf{z}, & (x,t)\in\Omega\times(0,+\infty), \\
\frac{\partial \mathbf{z}}{\partial {\eta}}=0,& (x,t)\in\partial\Omega\times(0,+\infty),\\
  \mathbf{z}(x,0)\geq 0, &x\in\Omega,
\end{array}
\right.
\end{equation}
 Then the solution of
(\ref{eq:lemma-conservation-3}) will converge to a finite limit as
time tends to infinity.
\end{lemma}

%
%
%
%

\subsection{Proof of  Theorem~\ref{thm:main-convergence}}

As Proposition~\ref{lemma:equibrium} indicates, the system has a unique constant equilibrium, namely $
  \mathbf{w}^{*}=\left(u^*_1, u^*_2, v^*_1,\cdots, v^*_4\right).$
The solution to system~(\ref{eq:ChFA-Model-XY-PDEs})-(\ref{eq:PDEs-Initial}) is denoted by
$$
    \mathbf{w}(x,t)=\left(u_1(x,t), u_2(x,t), v_1(x,t),\cdots, v_4(x,t)\right).
$$
The remainder work is to prove   $\mathbf{w}(x,t)\rightrightarrows \mathbf{w}^{*}, t\rightarrow \infty$.

First, by simple calculation, it is easy to obtain
\begin{proposition}
If $\mathbf{w}(x,t)$ is  the solution of system
(\ref{eq:ChFA-Model-XY-PDEs}) with  boundary condition (\ref{eq:PDEs-Boundary}) and initial condition
(\ref{eq:PDEs-Initial}), then
$$
    \mathbf{w}(x,t)+\left(
    \frac{1 }{a_1},   \frac{1}{a_2},  \frac{1}{a_1},  \frac{1+\frac{c_1}{a_1}}{c_2},
    \frac{1+\frac{c_4}{a_2}}{c_3},  \frac{1}{a_2}
    \right)\delta
$$
is the solution of system
(\ref{eq:ChFA-Model-XY-PDEs}) with   boundary condition (\ref{eq:PDEs-Boundary}) and the following initial condition
 \begin{eqnarray}\label{eq:new-initial-conditions}
\left(\phi_1(x),\phi_2(x),\psi_1(x),\psi_2(x),\psi_3(x),\psi_4(x)\right)
+\left(
    \frac{1 }{a_1},   \frac{1}{a_2},  \frac{1}{a_1},  \frac{1+\frac{c_1}{a_1}}{c_2},
    \frac{1+\frac{c_4}{a_2}}{c_3},  \frac{1}{a_2}
    \right)\delta.
 \end{eqnarray}
Here $\delta$ is a  constant.
\end{proposition}

\par According to this proposition, we may assume   solution $\mathbf{w}(x,t)$ of system~(\ref{eq:ChFA-Model-XY-PDEs})-(\ref{eq:PDEs-Initial}) has a positive lower bound. Otherwise, we  instead consider  new initial condition  (\ref{eq:new-initial-conditions})
and finally let positive $\delta$ tend to zero.

\par In the following we will define a sequence of lower and upper solutions. Scaling the equilibrium, we let
$$
   \underline{\mathbf{w}}^{(1)}(x,t)
   = (\underline{u}^{(1)}_1(x,t),\underline{u}^{(1)}_2(x,t),\underline{v}^{(1)}_1(x,t),\cdots,\underline{v}^{(1)}_4(x,t))= K_1  \mathbf{w}^{*}=\left(K_1u^*_1, K_1u^*_2,K_1v^*_1,\cdots, K_1v^*_4\right), $$
$$
 \bar{\mathbf{w}}^{(1)}(x,t)=(\bar{u}^{(1)}_1(x,t),\bar{u}^{(1)}_2(x,t),\bar{v}^{(1)}_1(x,t),\cdots,\bar{v}^{(1)}_4(x,t))=K_2  \mathbf{w}^{*}=\left(K_2u^*_1, K_2u^*_2,K_2v^*_1,\cdots, K_2v^*_4\right),
$$
where $0<K_1< K_2$ are  constant factors.
 Because the solution has a positive lower bound as mentioned above, and is bounded according to Theorem~\ref{thm:bounded-solution},
 we can choose appropriate  $K_1$ and $K_2$  such that
$$
    \underline{\mathbf{w}}^{(1)}(x,t)\leq \mathbf{w}(x,t)\leq \bar{\mathbf{w}}^{(1)}(x,t).
$$
That is, $\underline{\mathbf{w}}^{(1)}(x,t)$ and $\bar{\mathbf{w}}^{(1)}(x,t)$ are a pair of coupled lower and upper solutions of the system.

According to Lemma~\ref{lemma:lemma-conservation2}, $u_1(x,t)$ and $u_2(x,t)$ are rewritten as
\begin{equation}
u_1(x,t) =v_1(x,t)+v_2(x,t)-W_1+r_1(x,t),
\end{equation}
\begin{equation}
u_2(x,t) =v_3(x,t)+v_4(x,t)-W_2+r_2(x,t),
\end{equation}
where $r_i(x,t)\rightrightarrows 0, t\rightarrow \infty,$ for $i=1,2$.
Therefore, the following subsystem is essentially determined by $v_j(x,t)(j=1,2,\cdots,4)$:
\begin{equation}\label{eq:V-system}
\left\{
\begin{aligned}
\frac{\partial v_1}{\partial t}-\Delta v_1 &=-a_1\min\{u_1,v_1\}-c_1v_1+c_2v_2,\\
\frac{\partial v_2}{\partial t}-\Delta v_2 &=-c_2v_2+a_2\min\{u_2,v_4\}+c_1v_1,\\
\frac{\partial v_3}{\partial t}-\Delta v_3 &=-c_3v_3+c_4v_4+a_1\min\{u_1,v_1\},\\
\frac{\partial v_4}{\partial t}-\Delta v_4
&=-a_2\min\{u_2,v_4\}-c_4v_4+c_3v_3.
\end{aligned}
 \right.
\end{equation}
For $m\geq 1$, we define iterations:
\begin{eqnarray}\label{eq:V-iteration-1}
\left\{
\begin{split}
   &\frac{\partial \underline{v}^{(m+1)}_1}{\partial t}-\Delta \underline{v}^{(m+1)}_1
   =-a_1\min\{\bar{u}^{(m)}_1,\underline{v}^{(m+1)}_1\}-c_1\underline{v}^{(m+1)}_1+c_2\underline{v}^{(m)}_2,\\
   &\frac{\partial \underline{v}^{(m+1)}_2}{\partial t}-\Delta \underline{v}^{(m+1)}_2
   =-c_2\underline{v}^{(m+1)}_2+c_1\underline{v}^{(m)}_1+a_2\min\{\underline{u}^{(m)}_2,\underline{v}^{(m)}_4\},\\
  &\frac{\partial \underline{v}^{(m+1)}_3}{\partial t}-\Delta \underline{v}^{(m+1)}_3
   =-c_3\underline{v}^{(m+1)}_3+c_4\underline{v}^{(m)}_4+a_1\min\{\underline{u}^{(m)}_1,\underline{v}^{(m)}_1\},\\
  &\frac{\partial \underline{v}^{(m+1)}_4}{\partial t}-\Delta \underline{v}^{(m+1)}_4
   =-a_2\min\{\bar{u}^{(m)}_2,\underline{v}^{(m+1)}_4\} +c_3\underline{v}^{(m)}_3-c_4\underline{v}^{(m+1)}_4,
\end{split}\right.
\end{eqnarray}
and
\begin{eqnarray}\label{eq:V-iteration-2}
\left\{
\begin{split}
   &\frac{\partial \bar{v}^{(m+1)}_1}{\partial t}-\Delta \bar{v}^{(m+1)}_1
   =-a_1\min\{\underline{u}^{(m)}_1,\bar{v}^{(m+1)}_1\}-c_1\bar{v}^{(m+1)}_1+c_2\bar{v}^{(m)}_2,\\
   &\frac{\partial \bar{v}^{(m+1)}_2}{\partial t}-\Delta \bar{v}^{(m+1)}_2
   =-c_2\bar{v}^{(m+1)}_2+c_1\bar{v}^{(m)}_1+a_2\min\{\bar{u}^{(m)}_2,\bar{v}^{(m)}_4\},\\
  &\frac{\partial \bar{v}^{(m+1)}_3}{\partial t}-\Delta \bar{v}^{(m+1)}_3
   =-c_3\bar{v}^{(m+1)}_3+c_4\bar{v}^{(m)}_4+a_1\min\{\bar{u}^{(m)}_1,\bar{v}^{(m)}_1\},\\
  &\frac{\partial \bar{v}^{(m+1)}_4}{\partial t}-\Delta \bar{v}^{(m+1)}_4
   =-a_2\min\{\underline{u}^{(m)}_2,\bar{v}^{(m+1)}_4\} +c_3\bar{v}^{(m)}_3-c_4\bar{v}^{(m+1)}_4.
\end{split}\right.
\end{eqnarray}

%
%
%
%

In addition, they satisfy the  boundary   and initial conditions:
\begin{equation}\label{eq:V-Boundary}
        \frac{\partial
       \underline{v}^{(m+1)}_j}{\partial\eta}=\frac{\partial \bar{v}^{(m+1)}_j}{\partial\eta
       }=0, \ j=1,2,3,4.
\end{equation}
\begin{equation}\label{eq:V-Initial}
 \underline{v}^{(m+1)}_j(x,0)= \bar{v}^{(m+1)}_j(x,0)=\psi_j(x), \ j=1,2,3,4.
\end{equation}
In the above~(\ref{eq:V-iteration-1}) and~(\ref{eq:V-iteration-2}),  $\underline{u}^{(m)}_i, \bar{u}^{(m)}_i, i=1,2, m\geq 2, $ are defined as
\begin{eqnarray}\label{eq:V-u-definition}
\left\{
\begin{split}
   &\underline{u}^{(m)}_1=\underline{v}^{(m)}_1+\underline{v}^{(m)}_2-W_1+r_1(x,t),\\
   &\underline{u}^{(m)}_2=\underline{v}^{(m)}_3+\underline{v}^{(m)}_4-W_2+r_2(x,t),\\
  & \bar{u}^{(m)}_1=\bar{v}^{(m)}_1+\bar{v}^{(m)}_2-W_1+r_1(x,t),\\
  &\bar{u}^{(m)}_2=\bar{v}^{(m)}_3+\bar{v}^{(m)}_4-W_2+r_2(x,t).
\end{split}\right.
\end{eqnarray}
For $m\geq 1$, we denote
$$
  \underline{\mathbf{w}}^{(m)}(x,t)
  =\left(\underline{u}^{(m)}_1(x,t), \underline{u}^{(m)}_2(x,t),
  \underline{v}^{(m)}_1(x,t),\underline{v}^{(m)}_2(x,t),\cdots,\underline{v}^{(m)}_4(x,t)\right),
$$
$$
  \bar{\mathbf{w}}^{(m)}(x,t)
  =\left(\bar{u}^{(m)}_1(x,t), \bar{u}^{(m)}_2(x,t),
  \bar{v}^{(m)}_1(x,t),\bar{v}^{(m)}_2(x,t),\cdots,\bar{v}^{(m)}_4(x,t)\right).
$$
They satisfy
\begin{proposition} $\{\underline{\mathbf{w}}^{(m)}(x,t)\}_{m=1}^{\infty}$ and $\{\bar{\mathbf{w}}^{(m)}(x,t)\}_{m=1}^{\infty}$
are sequences of lower and upper solutions of system~(\ref{eq:ChFA-Model-XY-PDEs})-(\ref{eq:PDEs-Initial}), satisfying
$$
\underline{\mathbf{w}}^{(m)}(x,t)\leq \underline{\mathbf{w}}^{(m+1)}(x,t)\leq  {\mathbf{w}}(x,t) \leq \bar{\mathbf{w}}^{(m+1)}(x,t)\leq \bar{\mathbf{w}}^{(m)}(x,t).
$$
\end{proposition}
\begin{proof} We prove this proposition by induction. Clearly, when $m=1$, the conclusion holds.
Suppose the result holds for $m\geq 1$, i.e., $\underline{\mathbf{w}}^{(m)}$ and $\bar{\mathbf{w}}^{(m)}$ are lower and upper solutions
respectively. By (\ref{eq:V-iteration-1}) and (\ref{eq:V-iteration-2}),
\begin{eqnarray}
\begin{split}
   \frac{\partial \underline{v}^{(m+1)}_1}{\partial t}-\Delta \underline{v}^{(m+1)}_1
   &=-a_1\min\{\bar{u}^{(m)}_1,\underline{v}^{(m+1)}_1\}-c_1\underline{v}^{(m+1)}_1+c_2\underline{v}^{(m)}_2,\\
   &\leq -a_1\min\{u_1,\underline{v}^{(m+1)}_1\}-c_1\underline{v}^{(m+1)}_1+c_2v_2,
\end{split}
\end{eqnarray}
\begin{eqnarray}
\begin{split}
   \frac{\partial \bar{v}^{(m+1)}_1}{\partial t}-\Delta \bar{v}^{(m+1)}_1
   &=-a_1\min\{\underline{u}^{(m)}_1,\bar{v}^{(m+1)}_1\}-c_1\bar{v}^{(m+1)}_1+c_2\bar{v}^{(m)}_2,\\
   & \geq -a_1\min\{u_1,\bar{v}^{(m+1)}_1\}-c_1\bar{v}^{(m+1)}_1+c_2v_2.\\
\end{split}
\end{eqnarray}
That is, $\underline{v}^{(m+1)}_1$ and $\bar{v}^{(m+1)}_1$ are   lower and upper solutions of $v_1$ respectively. Similarly,
we can prove the case of $v_j, j=2,3,4$. As a consequence, $\underline{u}^{(m+1)}_i$ and  $\bar{u}^{(m+1)}_i$ defined through
equations (\ref{eq:V-u-definition}) are  lower and upper solutions of $u_i$ where $i=1,2$.

\par The monotone property can also be proved inductively. Clearly, by simple calculation,
$$\underline{\mathbf{w}}^{(1)}(x,t)\leq \underline{\mathbf{w}}^{(2)}(x,t);\quad \bar{\mathbf{w}}^{(2)}(x,t)\leq \bar{\mathbf{w}}^{(1)}(x,t).$$
Suppose that, for $m\geq2$,
$$\underline{\mathbf{w}}^{(m-1)}(x,t)\leq \underline{\mathbf{w}}^{(m)}(x,t),\quad \bar{\mathbf{w}}^{(m)}(x,t)\leq \bar{\mathbf{w}}^{(m-1)}(x,t).$$
We will prove that
 $$\underline{\mathbf{w}}^{(m)}(x,t)\leq \underline{\mathbf{w}}^{(m+1)}(x,t),\quad \bar{\mathbf{w}}^{(m+1)}(x,t)\leq \bar{\mathbf{w}}^{(m)}(x,t).$$
Notice that
\begin{eqnarray}
\begin{split}
\frac{\partial \underline{v}^{(m)}_1}{\partial t}-\Delta \underline{v}^{(m)}_1
   &=-a_1\min\{\bar{u}^{(m-1)}_1,\underline{v}^{(m)}_1\}-c_1\underline{v}^{(m)}_1+c_2\underline{v}^{(m-1)}_2,\\
   \frac{\partial \underline{v}^{(m+1)}_1}{\partial t}-\Delta \underline{v}^{(m+1)}_1
   &=-a_1\min\{\bar{u}^{(m)}_1,\underline{v}^{(m+1)}_1\}-c_1\underline{v}^{(m+1)}_1+c_2\underline{v}^{(m)}_2.
\end{split}
\end{eqnarray}
Because $\underline{v}^{(m-1)}_2\leq \underline{v}^{(m)}_2$ and $\bar{u}^{(m)}_1\leq \bar{u}^{(m-1)}_1$, so by   comparison principle, it is easy to see $\underline{v}^{(m)}_1\leq \underline{v}^{(m+1)}_1$. The proofs for other cases are similar, and thus omitted.
\end{proof}

\begin{proposition}
For any $m\geq 1$, $\underline{\mathbf{w}}^{(m)}(x,t)$ and $\bar{\mathbf{w}}^{(m)}(x,t)$ converge to constants   uniformly, as time tends to infinity.
\end{proposition}
\begin{proof}
We adopt an induction method. Clearly, when $m=1$, the conclusion holds because $\underline{\mathbf{w}}^{(1)}(x,t)$ and $\bar{\mathbf{w}}^{(1)}(x,t)$  are constants. Assume that for $m\geq 1$, there exist constant vectors
$\underline{\mathbf{k}}^{(m)}$ and $\bar{\mathbf{k}}^{(m)}$ such that
$$\underline{\mathbf{w}}^{(m)}(x,t)\rightrightarrows \underline{\mathbf{k}}^{(m)},\quad \bar{\mathbf{w}}^{(m)}(x,t)\rightrightarrows \bar{\mathbf{k}}^{(m)}, \quad t\rightarrow\infty.$$
So we can rewrite
$$
   \underline{w}^{(m)}_i(x,t)=\underline{k}^{(m)}_i+\underline{r}^{(m)}_i(x,t),
   \ \bar{w}^{(m)}_i(x,t)=\bar{k}^{(m)}_i+\bar{r}^{(m)}_i(x,t), \ i=1,2,\cdots,6,
$$
where $\underline{k}^{(m)}_i$ and $\bar{k}^{(m)}_i$ are constants and all $\underline{r}^{(m)}_i(x,t)$ and $\bar{r}^{(m)}_i(x,t)$ converge to zeros uniformly as time goes to infinity.
Therefore, we can write
\begin{eqnarray}
\begin{split}
\frac{\partial \underline{v}^{(m+1)}_1}{\partial t}-\Delta \underline{v}^{(m+1)}_1
     &=-a_1\min\{\bar{u}^{(m)}_1,\underline{v}^{(m+1)}_1\}-c_1\underline{v}^{(m+1)}_1+c_2\underline{v}^{(m)}_2\\
     &=-a_1\min\{\bar{k}^{(m)}_1,\underline{v}^{(m+1)}_1\}-c_1\underline{v}^{(m+1)}_1+c_2\underline{k}^{(m)}_4+ r(x,t),
\end{split}
\end{eqnarray}
where $r(x,t)\rightrightarrows 0$ as $t\rightarrow\infty$. Then by Lemma~\ref{lemma:convergence-2}, we know that
$\underline{v}^{(m+1)}_1(x,t)$ uniformly converges to a constant as time tends to infinity.  We can similarly prove the uniform
convergence for all
$\underline{w}^{(m)}_i(x,t)$ and $\bar{w}^{(m)}_i(x,t)$.
\end{proof}

From the above proof, we know
 $$
               \underline{\mathbf{k}}_{m}= \lim_{t\rightarrow\infty}\underline{\mathbf{w}}^{(m)}w(x,t)\leq \liminf_{t\rightarrow\infty} \mathbf{w}(x,t)
                \leq   \limsup_{t\rightarrow\infty}\mathbf{w}(x,t)\leq
                \lim_{t\rightarrow\infty}  \bar{\mathbf{w}}^{(m)} (x,t)=\bar{\mathbf{k}}_{m}.
$$
Because $\{\underline{\mathbf{k}}^{(m)}\}_{m=1}^{\infty}$ and $\{\bar{\mathbf{k}}^{(m)}\}_{m=1}^{\infty}$  are bounded monotone sequences,
we denote
$$
\lim_{m\rightarrow\infty}\underline{\mathbf{k}}^{(m)}= \underline{\mathbf{w}}^*=(\underline{u}_1^*,\underline{u}_2^*,\underline{v}_1^*,\underline{v}_2^*,\underline{v}_3^*,\underline{v}_4^*),
\quad
\lim_{m\rightarrow\infty}\bar{\mathbf{k}}^{(m)}= \bar{\mathbf{w}}^*=
(\bar{u}_1^*,\bar{u}_2^*,\bar{v}_1^*,\bar{v}_2^*,\bar{v}_3^*,\bar{v}_4^*).$$
Let $m\rightarrow\infty$, we   have
\begin{equation}
 \underline{\mathbf{w}}^*   \leq \liminf_{t\rightarrow\infty} \mathbf{w}(x,t)
                \leq   \limsup_{t\rightarrow\infty}\mathbf{w}(x,t)\leq  \bar{\mathbf{w}}^*.
\end{equation}
In the following, we will prove Theorem~\ref{thm:main-convergence}, i.e., the uniform convergence of $\mathbf{w}(x,t)$.

\begin{proof}
By (\ref{eq:V-iteration-1}), (\ref{eq:V-iteration-2}) and (\ref{eq:V-u-definition}),  $\underline{\mathbf{w}}^*$ and
$\bar{\mathbf{w}}^*$ satisfy that
 \begin{eqnarray}\label{eq:V-iteration-3}
\left\{
\begin{split}
  -a_1\min\{\bar{u}^{(*)}_1,\underline{v}^{(*)}_1\}-c_1\underline{v}^{(*)}_1+c_2\underline{v}^{(*)}_2=0,\\
    -c_2\underline{v}^{(*)}_2+c_1\underline{v}^{(*)}_1+a_2\min\{\underline{u}^{(*)}_2,\underline{v}^{(*)}_4\}=0,\\
  -c_3\underline{v}^{(*)}_3+c_4\underline{v}^{(*)}_4+a_1\min\{\underline{u}^{(*)}_1,\underline{v}^{(*)}_1\}=0,\\
   -a_2\min\{\bar{u}^{(*)}_2,\underline{v}^{(*)}_4\} +c_3\underline{v}^{(*)}_3-c_4\underline{v}^{(*)}_4=0,
\end{split}\right.
\end{eqnarray}
\begin{eqnarray}\label{eq:V-iteration-4}
\left\{
\begin{split}
   -a_1\min\{\underline{u}^{(*)}_1,\bar{v}^{(*)}_1\}-c_1\bar{v}^{(*)}_1+c_2\bar{v}^{(*)}_2=0,\\
   -c_2\bar{v}^{(*)}_2+c_1\bar{v}^{(*)}_1+a_2\min\{\bar{u}^{(*)}_2,\bar{v}^{(*)}_4\}=0,\\
 -c_3\bar{v}^{(*)}_3+c_4\bar{v}^{(*)}_4+a_1\min\{\bar{u}^{(*)}_1,\bar{v}^{(*)}_1\}=0,\\
 -a_2\min\{\underline{u}^{(*)}_2,\bar{v}^{(*)}_4\} +c_3\bar{v}^{(*)}_3-c_4\bar{v}^{(*)}_4=0,
\end{split}\right.
\end{eqnarray}
\begin{eqnarray}\label{eq:V-u-5}
\left\{
\begin{split}
   &\underline{u}^{(*)}_1=\underline{v}^{(*)}_1+\underline{v}^{(*)}_2-W_1,\\
   &\underline{u}^{(*)}_2=\underline{v}^{(*)}_3+\underline{v}^{(*)}_4-W_2,\\
  & \bar{u}^{(*)}_1=\bar{v}^{(*)}_1+\bar{v}^{(*)}_2-W_1,\\
  &\bar{u}^{(*)}_2=\bar{v}^{(*)}_3+\bar{v}^{(*)}_4-W_2.
\end{split}\right.
\end{eqnarray}
According to (\ref{eq:V-iteration-3}),
$$
a_1\min\{\bar{u}^{(*)}_1,\underline{v}^{(*)}_1\}=a_2\min\{\underline{u}^{(*)}_2,\underline{v}^{(*)}_4\},  \quad
a_1\min\{\underline{u}^{(*)}_1,\underline{v}^{(*)}_1\}=a_2\min\{\bar{u}^{(*)}_2,\underline{v}^{(*)}_4\}.
$$
By $a_2\min\{\underline{u}^{(*)}_2, \underline{v}^{(*)}_4\} \leq a_2\min\{\bar{u}^{(*)}_2,\underline{v}^{(*)}_4\}$,
we know that  $a_1\min\{\bar{u}^{(*)}_1,\underline{v}^{(*)}_1\}\leq a_1\min\{\underline{u}^{(*)}_1,\underline{v}^{(*)}_1\}$
and hence,
$a_1\min\{\bar{u}^{(*)}_1,\underline{v}^{(*)}_1\}=a_1\min\{\underline{u}^{(*)}_1,\underline{v}^{(*)}_1\}.$
Consequently,
\begin{equation}\label{eq:V-iteration-5}
a_1\min\{\bar{u}^{(*)}_1,\underline{v}^{(*)}_1\}=a_2\min\{\underline{u}^{(*)}_2,\underline{v}^{(*)}_4\}=
  a_2\min\{\bar{u}^{(*)}_2,\underline{v}^{(*)}_4\}= a_1\min\{\underline{u}^{(*)}_1,\underline{v}^{(*)}_1\}.
\end{equation}
Similarly, from (\ref{eq:V-iteration-4}), we have that
\begin{equation}\label{eq:V-iteration-6}
a_1\min\{\underline{u}^{(*)}_1,\bar{v}^{(*)}_1\}=a_2\min\{\bar{u}^{(*)}_2,\bar{v}^{(*)}_4\}=
a_1\min\{\bar{u}^{(*)}_1,\bar{v}^{(*)}_1\}=a_2\min\{\underline{u}^{(*)}_2,\bar{v}^{(*)}_4\}.
\end{equation}


In (\ref{eq:V-iteration-6}),  $\min\{\underline{u}^{(*)}_1,\bar{v}^{(*)}_1\}= \min\{\bar{u}^{(*)}_1,\bar{v}^{(*)}_1\}$ means
that either $\bar{u}^{(*)}_1=\underline{u}^{(*)}_1$ or $\bar{v}^{(*)}_1\leq \underline{u}^{(*)}_1$.  Similarly,
$\min\{\bar{u}^{(*)}_2,\bar{v}^{(*)}_4\}= \min\{\underline{u}^{(*)}_2,\bar{v}^{(*)}_4\}$
implies that
$\bar{u}^{(*)}_2=\underline{u}^{(*)}_2$ or $\bar{v}^{(*)}_4\leq \underline{u}^{(*)}_2$. We divide into three cases to discuss.

\par (i) If  $\bar{u}^{(*)}_1=\underline{u}^{(*)}_1$,
 then $\underline{v}^{(*)}_1+\underline{v}^{(*)}_2=\bar{v}^{(*)}_1+\bar{v}^{(*)}_2$, according to (\ref{eq:V-u-5}). Since $\underline{v}^{(*)}_i\leq \bar{v}^{(*)}_i, i=1,2$,
 we have that $\underline{v}^{(*)}_i=\bar{v}^{(*)}_i, i=1,2$.
 By (\ref{eq:V-iteration-3}) and (\ref{eq:V-iteration-4}), we know
 $$
 -c_3\underline{v}^{(*)}_3+c_4\underline{v}^{(*)}_4=-c_3\bar{v}^{(*)}_3+c_4\bar{v}^{(*)}_4.
 $$
This implies that   $\underline{v}^{(*)}_i=\bar{v}^{(*)}_i, i=3,4$, and further leads to $\underline{u}^{(*)}_2=\bar{u}^{(*)}_2$ by (\ref{eq:V-u-5}).
So we have $\underline{\mathbf{w}}^* =\bar{\mathbf{w}}^*$, implying $\lim_{t\rightarrow\infty} \mathbf{w}(x,t)=\underline{\mathbf{w}}^* =\bar{\mathbf{w}}^*$.

\par (ii) If $\bar{u}^{(*)}_2=\underline{u}^{(*)}_2$, we still have $\underline{\mathbf{w}}^* =\bar{\mathbf{w}}^*$ and the discussion is similar to case (i).


\par (iii) If
$\bar{v}^{(*)}_1\leq \underline{u}^{(*)}_1$ and $\bar{v}^{(*)}_4\leq \underline{u}^{(*)}_2$,
then  it is clear to see
$$
   \limsup_{t\rightarrow\infty}v_1(x,t)\leq \bar{v}^{(*)}_1\leq \underline{u}^{(*)}_1\leq \liminf_{t\rightarrow\infty}u_1(x,t),
$$
$$
   \limsup_{t\rightarrow\infty}v_4(x,t) \leq \bar{v}^{(*)}_4\leq \underline{u}^{(*)}_2 \leq \liminf_{t\rightarrow\infty}u_2(x,t).
$$
So there exists a time $T_1>0$ such that after time $T_1$,
$$
  \min\{u_1,v_1\}=v_1,\quad \min\{u_2,v_4\}=v_4,
$$
and therefore, the subsystem (\ref{eq:V-system}) consists of $v_j$
becomes linear after time $T_1$:
\begin{equation}\label{eq:V-system-Q4}
\left\{
\begin{aligned}
\frac{\partial v_1}{\partial t}-\Delta v_1 &=-a_1v_1-c_1v_1+c_2v_2,\\
\frac{\partial v_2}{\partial t}-\Delta v_2 &=-c_2v_2+a_2v_4+c_1v_1,\\
\frac{\partial v_3}{\partial t}-\Delta v_3 &=-c_3v_3+c_4v_4+a_1v_1,\\
\frac{\partial v_4}{\partial t}-\Delta v_4 &=-a_2v_4-c_4v_4+c_3v_3.
\end{aligned}
 \right.
\end{equation}
According to   Lemma~\ref{lemma:lemma-conservation3},
$v_j(j=1,2,\cdots,4)$ converge to constants uniformly. As a result, $u_i(i=1,2)$ also
uniformly converge to limits. The proof of Theorem~\ref{thm:main-convergence} is completed.

\end{proof}

\begin{figure}[htbp]
\begin{center}
  \includegraphics[width=7.4cm]{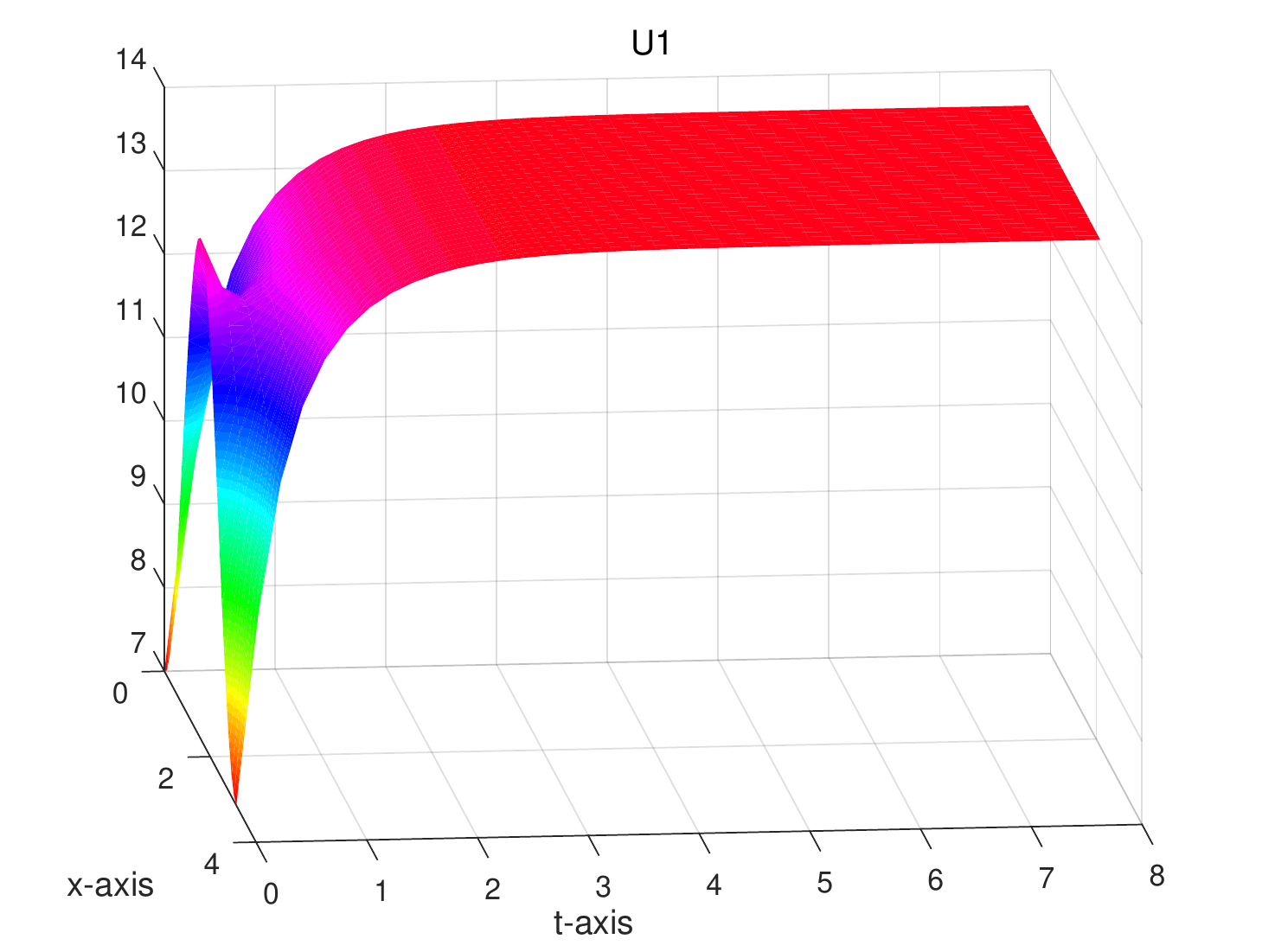}
 \includegraphics[width=7.4cm]{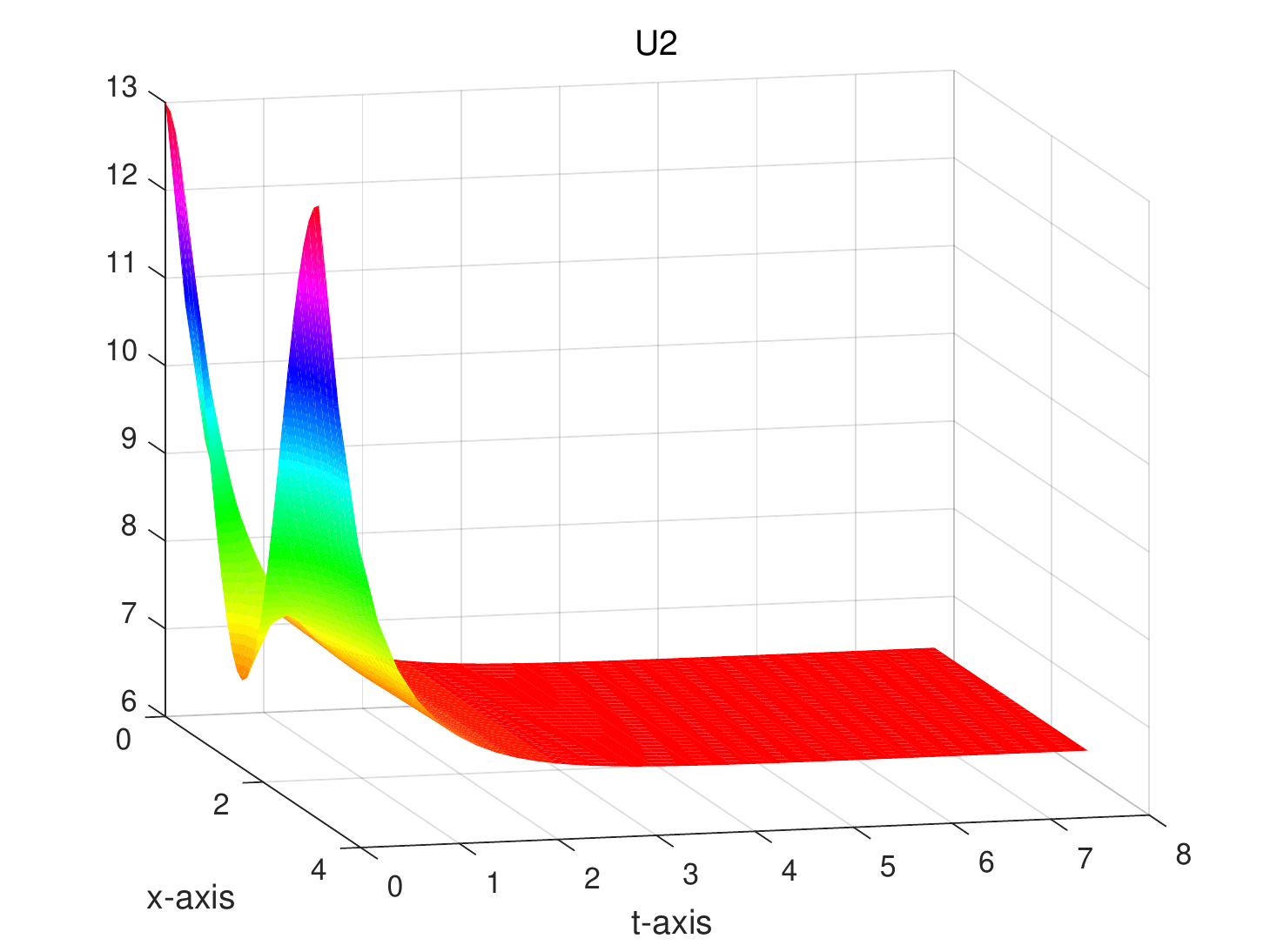}\\
 \includegraphics[width=7.4cm]{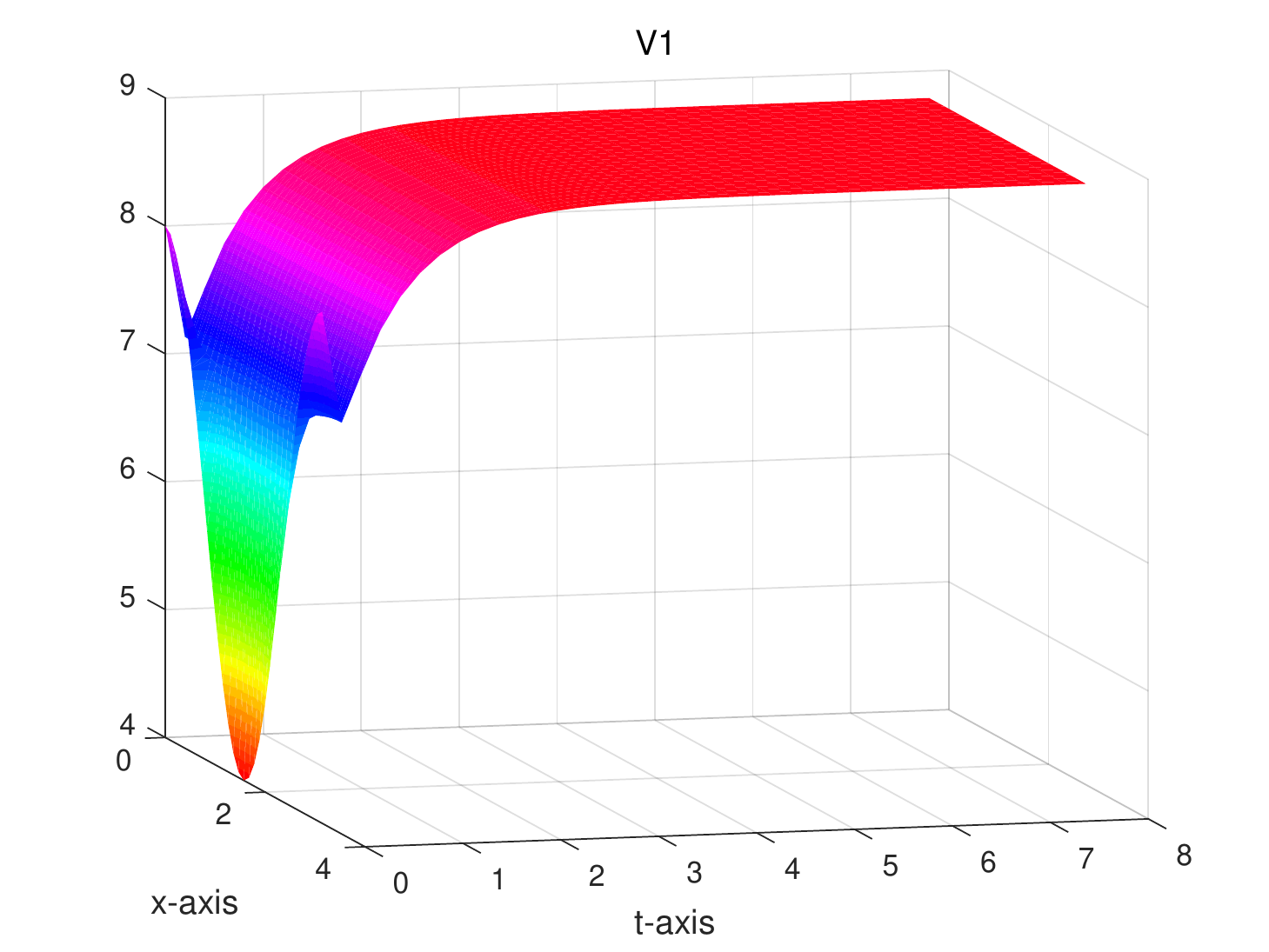}
 \includegraphics[width=7.4cm]{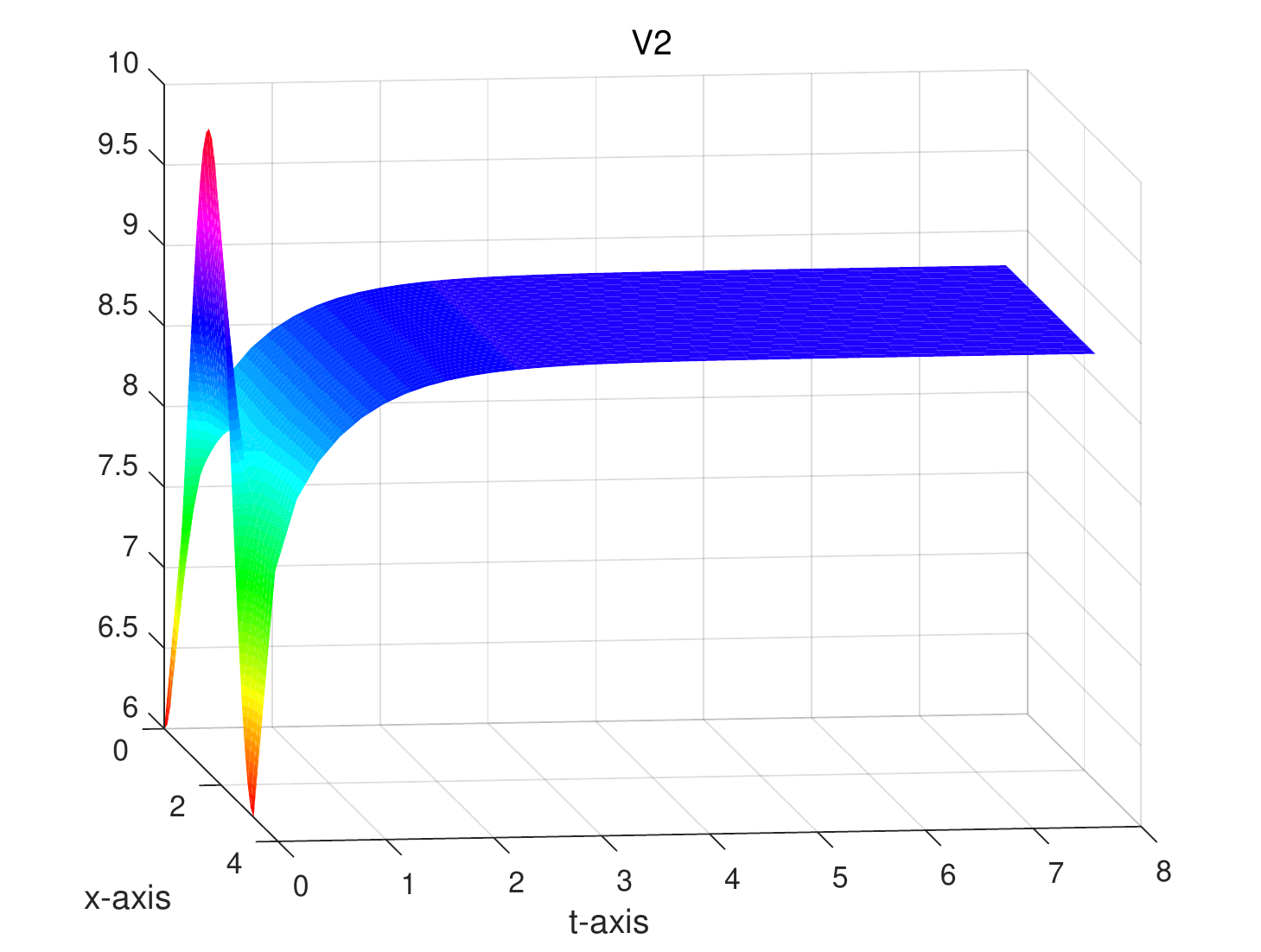}\\
 \includegraphics[width=7.4cm]{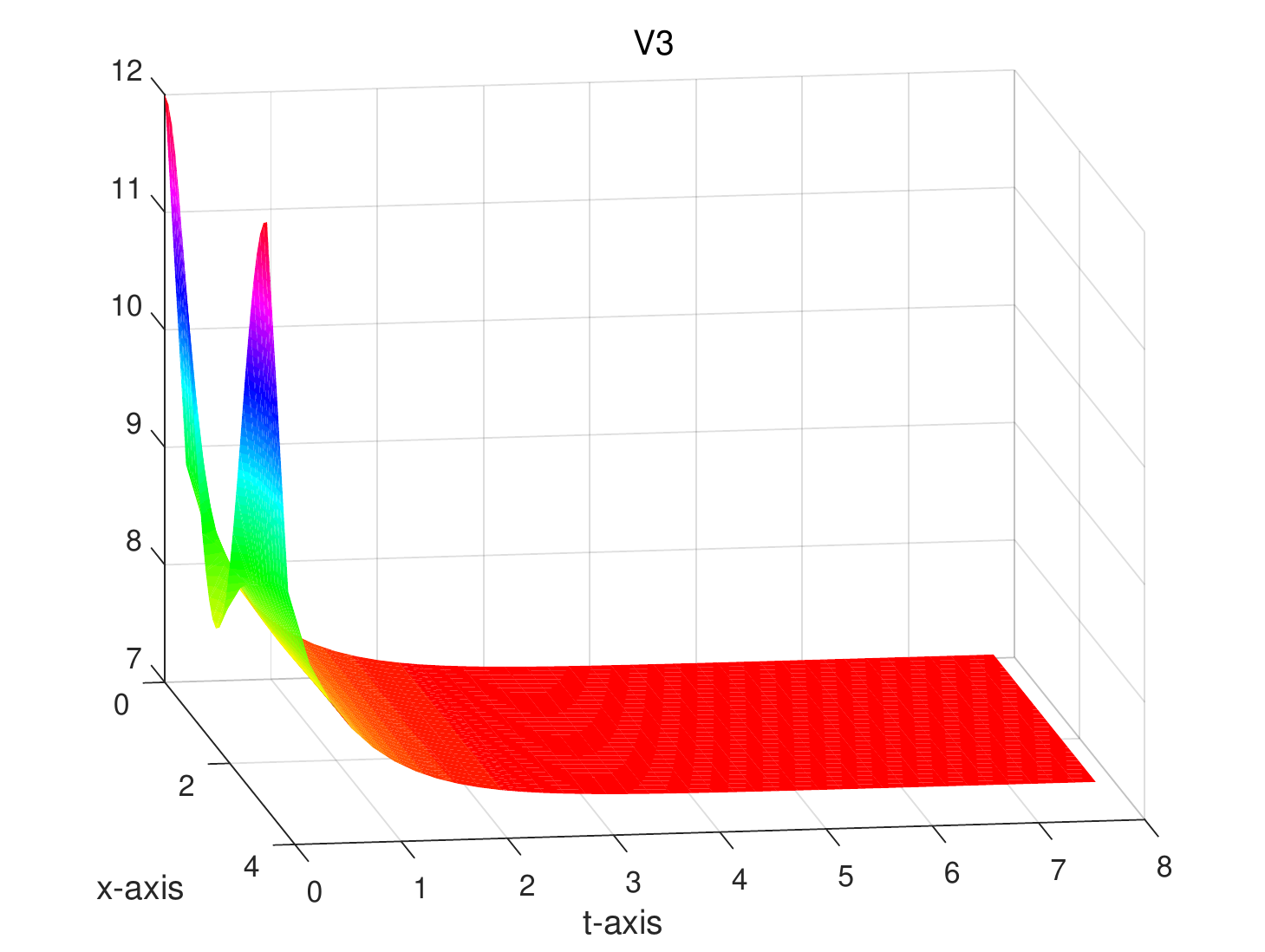}
  \includegraphics[width=7.4cm]{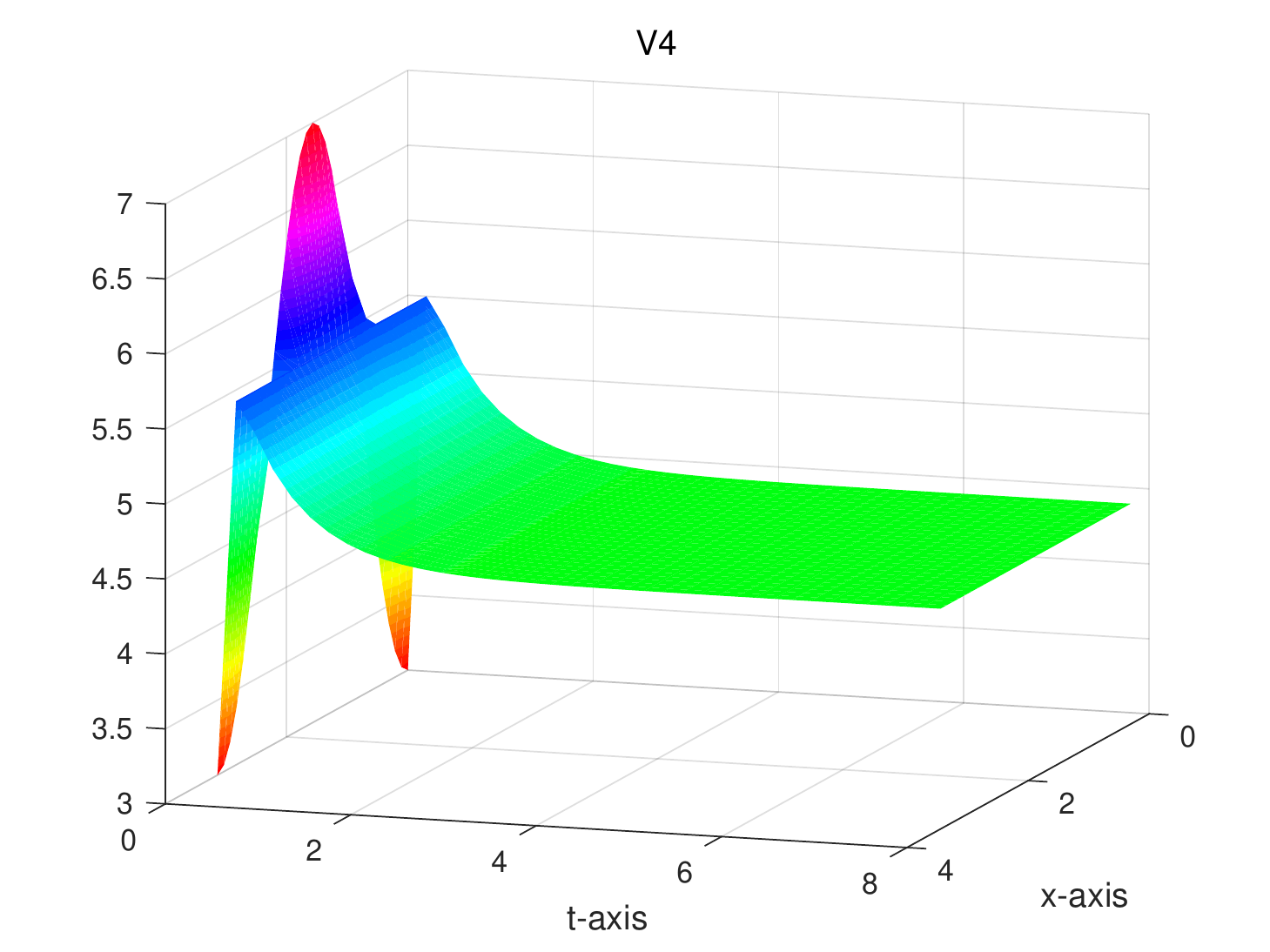}
  \caption{$Q_4$ finally dominates with the equilibrium $( 13.5757,\  6.4243,\   8.7879,\  8.7879,\ 7.0303,\ 4.3940)$
  when $(a_1,\ a_2,\ c_1,\ c_2,\ c_3,\ c_4)=(1,\ 2,\ 3,\ 4,\ 5,\ 6)$ }
  \label{fig:Num-solution-Q4}
  \end{center}
\end{figure}

\section{Numerical experiments}

This section presents some numerical experiments for  system~(\ref{eq:ChFA-Model-XY-PDEs})-(\ref{eq:PDEs-Initial}).
The parameters are set as
$$(a_1,a_2,c_1,c_2,c_3,c_4)=(1,2,3,4,5,6).$$
The considered region is
$\Omega=[0,\pi]$. The initial conditions   are   assumed as
$$
   \phi_1=10-3\cos(2x), \quad \phi_2=10+3\cos(2x),$$
   $$\psi_1=6+2\cos(2x), \quad \psi_2=8-2\cos(2x), \quad \psi_2=10+2\cos(2x), \quad \psi_2=5-2\cos(2x),
$$
where $x\in \Omega$. Therefore,
$$M_0=\frac{1}{\pi}\int_{0}^{\pi}(\phi_1+\phi_2)=20, \quad
N_0=\frac{1}{\pi}\int_{0}^{\pi}(\psi_1+\psi_2+\psi_3+\psi_4)=29,$$
and
$$W_1=\frac{1}{\pi}\int_{0}^{\pi}(\psi_1+\psi_2-\phi_1)=4, \quad
W_2=\frac{1}{\pi}\int_{0}^{\pi}(\psi_3+\psi_4-\phi_2)=5.$$

By simple calculation, $D_1=a_1c_2(a_2+c_4)+a_2c_3(a_1+c_1)+c_2c_3(a_1+a_2)=132$, and condition $(I_3^c)\wedge I_4^c$ holds because
\begin{equation*}\label{eq:I3}
  (I_3^c): a_2c_3(a_1 + c_1)N_0=1160 \geq 528= D_1W_1,
\end{equation*}
\begin{equation*}\label{eq:I4}
  (I_4^c): a_1c_2(a_2 + c_4)N_0=928 \geq 660= D_1W_2.
\end{equation*}
According to Lemma~\ref{lemma:equibrium}, $Q_4$  dominates the system and  the
equilibrium point   is
$$( 13.5757, 6.4243,   8.7879,   8.7879, 7.0303,  4.3940).$$
The numerical solution  of system~(\ref{eq:ChFA-Model-XY-PDEs})-(\ref{eq:PDEs-Initial}) is illustrated in Figure~\ref{fig:Num-solution-Q4}, which is consistent with the
uniform convergence result stated in  Theorem~\ref{thm:main-convergence}.

\section{Conclusions}
This paper has demonstrated a lower and upper solution method to
investigate the asymptotic behaviour of the conservative
reaction-diffusion system which is associated with a Markovian process algebra model.
In particular, we have proved the uniform convergence of the solution with
time to its constant equilibrium for a case study, together with experimental results
illustrations. As future work, the techniques and results established in this paper
are expected to extend to more general Markov process algebra models.
In addition, the relationship among the reaction-diffusions, the fluid approximations
and the Markov chains, will be further investigated in the future.

\bibliographystyle{elsarticle-num}
\bibliography{AMM2022Arxiv}

\end{document}